\documentclass[epj,referee]{svjour}

\usepackage{cite}
\usepackage{graphicx}

\begin{document}

\title{Comment on: ``Quantum aspects of the Lorentz symmetry violation on an
electron in a nonuniform electric field'' Eur. Phys. J. Plus (2020) 135:623}

\author{Paolo Amore \inst{1} \and Francisco M. Fern\'andez \inst{2}}

\institute{Facultad de Ciencias, CUICBAS, Universidad de Colima,
Bernal D\'{\i}az del Castillo 340, Colima, Colima, Mexico
\email{paolo@ucol.mx} \and INIFTA, Divisi\'{o}n Qu\'{\i}mica
Te\'{o}rica, Blvd. 113 y 64 (S/N), Sucursal 4, Casilla de Correo
16, 1900 La Plata, Argentina \email{fernande@quimica.unlp.edu.ar}}

\date{}

\abstract{ We analyze recent results concerning the hypothesis of
a privileged direction in the space-time that is made by
considering a background of the Lorentz symmetry violation
determined by a fixed spacelike vector field and the analysis of
quantum effects of this background on the interaction of a
nonrelativistic electron with a nonuniform electric field produced
by a uniform electric charge distribution. We show that the
conclusions derived by the authors are an artifact of the
truncation of the Frobenius series by means of the tree-term
recurrence relation for the expansion coefficients. Thus, the
existence of allowed angular frequencies stemming from this
procedure is meaningless and unphysical.}

\titlerunning{Misinterpretation of a quasi-exactly solvable model}

\maketitle

\onecolumn

In a recent paper Oliveira et al\cite{OBB20} consider the hypothesis of a
privileged direction in the space-time. To this end they take into account a
background of the Lorentz symmetry violation determined by a fixed spacelike
vector field and analyze quantum effects of this background on the
interaction of a nonrelativistic electron with a nonuniform electric field
produced by a uniform electric charge distribution. Thus, they derive a
Schr\"{o}dinger equation that is separable in cylindrical coordinates. The
resulting radial eigenvalue equation, which exhibits Coulomb plus harmonic
interactions, is treated by means of the Frobenius method. Since the
expansion coefficients satisfy a three-term recurrence relation the authors
can force a truncation of the power series to obtain polynomial solutions
and exact eigenvalues. From such results the authors predict the existence
of allowed angular frequencies. In what follows we discuss the effect of the
truncation of the Frobenius series on the physical conclusions drawn by the
authors.

The starting point of present discussion is the eigenvalue equation for the
dimensionless radial function
\begin{eqnarray}
&&u^{\prime \prime }(x)+\frac{1}{x}u^{\prime }(x)-\frac{\nu ^{2}}{x^{2}}%
u(x)-\delta xu(x)-x^{2}u(x)+Wu(x)=0,  \nonumber \\
&&W=\frac{2\tau }{m\omega },\;\tau =2m\mathcal{E}-k^{2},\;\omega =\sqrt{%
\frac{2|q|\rho }{m}},\;\nu =l+\frac{1}{2}(1-s),  \nonumber \\
&&\delta=\frac{gb\sqrt{2m\omega }}{|q|},  \label{eq:eigen_eq}
\end{eqnarray}
where $l=0,\pm 1,\pm 2,\ldots $ is the rotational quantum number, $s=\pm 1$,
$m$ the mass of the particle, $q=-|q|$ an electric charge, $b>0$ the
magnitude of a fixed spaced-like vector field, $\rho >0$ is related to the
uniform electric charge distribution and $\mathcal{E}$ the energy. The
constant $-\infty <k<\infty $ comes from the fact that the motion is
unbounded along the $z$ axis; therefore the spectrum is continuous an
bounded from below $\mathcal{E}\geq \frac{\tau }{2m}$. The authors simply
set $\hbar =1$, $c=1$ though there are well known procedures for obtaining
suitable dimensionless equations in a clearer and more rigorous way\cite{F20}%
. In what follows we focus on the discrete values of $W$ that one obtains
from the bound-state solutions of equation (\ref{eq:eigen_eq}) that satisfy
\begin{equation}
\int_{0}^{\infty }\left| u(x)\right| ^{2}x\,dx<\infty .
\label{eq:bound_states}
\end{equation}
Notice that we have bound states for all $-\infty <\delta <\infty $ and that
the eigenvalues $W$ satisfy
\begin{equation}
\frac{\partial W}{\partial \delta }=\left\langle x\right\rangle >0,
\label{eq:HFT}
\end{equation}
according to the Hellmann-Feynman theorem\cite{F39}.

The eigenvalue equation (\ref{eq:eigen_eq}) is an example of conditionally
solvable (or quasi-exactly solvable) problems that have been widely studied
by several authors and exhibit a hidden algebraic structure (see\cite{T16}
and references therein).

In order to solve the eigenvalue equation (\ref{eq:eigen_eq}) the authors
propose the ansatz
\begin{equation}
u(x)=x^{\gamma }\exp \left( -\frac{\delta }{2}x-\frac{x^{2}}{2}\right)
f(x),\;f(x)=\sum_{j=0}^{\infty }a_{j}x^{j},\;\gamma =|\nu |,
\label{eq:ansatz}
\end{equation}
and derive the three-term recurrence relation
\begin{eqnarray}
a_{j+2} &=&\frac{\delta \left( 2j+2\gamma +3\right) }{2\left( j+2\right)
\left[ j+2\left( \gamma +1\right) \right] }a_{{j+1}}+\frac{2j-\Theta }{%
\left( j+2\right) \left[ j+2\left( \gamma +1\right) \right] }a_{{j}},
\nonumber \\
\Theta &=&W-2(\gamma +1)+\frac{\delta ^{2}}{4},\;j=-1,0,\ldots
,\;a_{-1}=0,\;a_{0}=1.  \label{eq:TTRR}
\end{eqnarray}
If the truncation condition $a_{n+1}=a_{n+2}=0$ has physically acceptable
solutions then one obtains exact eigenvalues and eigenfunctions. The reason
is that $a_{j}=0$ for all $j>n$ and the factor $f(x)$ in equation (\ref
{eq:ansatz}) reduces to a polynomial of degree $n$. This truncation
condition is equivalent to $\Theta =2n$ and $a_{n+1}=0$. The latter equation
is a polynomial function of $\delta $ of degree $n+1$ and it can be proved
that all the roots $\delta _{\gamma }^{(n,i)}$, $i=1,2,\ldots ,n+1$, are real%
\cite{CDW00,AF20}. If $V(\delta ,x)=\delta x+x^{2}$ denotes the
parameter-dependent potential for the model discussed here, then it is clear
that the truncation condition produces eigenvalues $W_{\gamma
}^{(n,i)}=2(n+\gamma +1)-\frac{\left[ \delta _{\gamma }^{(n,i)}\right] ^{2}}{%
4}$ for $n+1$ different potential-energy functions $V_{\gamma
}^{(n,i)}(x)=V\left( \delta _{\gamma }^{(n,i)},x\right) $. It is worth
noticing that the truncation condition only yields \textit{some particular}
eigenvalues and eigenfunctions because not all the solutions $u(x)$
satisfying equation (\ref{eq:bound_states}) have polynomial factors $f(x)$.
From now on we will refer to them as follows
\begin{equation}
u_{\gamma }^{(n,i)}(x)=x^{\gamma }f_{\gamma }^{(n,i)}(y)\exp \left( -\frac{%
\delta }{2}x-\frac{x^{2}}{2}\right) ,\;f_{\gamma
}^{(n,i)}(x)=\sum_{j=0}^{n}a_{j,\gamma }^{(n,i)}x^{j}.  \label{eq:f^(n,i)(y)}
\end{equation}

Let us discuss the first cases as illustrative examples. When $n=0$ we have $%
\delta _{\gamma }^{(0)}=0$ and the eigenfunction $u_{\gamma }^{(0)}(x)$ has
no nodes. We may consider this case trivial because the problem reduces to
the exactly solvable harmonic oscillator. Probably for this reason it was
not explicitly taken into account by Oliveira et al\cite{OBB20}.

When $n=1$ there are two roots
\begin{equation}
\delta _{\gamma }^{(1,1)}=\frac{2\sqrt{2}}{\sqrt{2\gamma +3}},\;\delta
_{\gamma }^{(1,2)}=-\frac{2\sqrt{2}}{\sqrt{2\gamma +3}},
\label{eq:delta_n=1}
\end{equation}
with the corresponding coefficients
\begin{equation}
a_{1,\gamma }^{(1,1)}=\frac{\sqrt{2}}{\sqrt{2\gamma +3}},\;a_{1,\gamma
}^{(1,2)}-\frac{\sqrt{2}}{\sqrt{2\gamma +3}},  \label{eq:a_j_n=1}
\end{equation}
respectively. We appreciate that the eigenfunction $u_{\gamma }^{(1,1)}(x)$
is nodeless and $u_{\gamma }^{(1,2)}(x)$ has one node.

For $n=2$ the results are
\begin{eqnarray}
\delta _{\gamma }^{(2,1)} &=&4\sqrt{\frac{4\gamma +7}{\left( 2\gamma
+3\right) \left( 2\gamma +5\right) }},\;a_{1,\gamma }^{(2,1)}=2\sqrt{\frac{%
4\gamma +7}{\left( 2\gamma +3\right) \left( 2\gamma +5\right) }},\;
\nonumber \\
a_{2,\gamma }^{(2,1)} &=&\frac{2}{2\gamma +5},  \nonumber \\
\delta _{\gamma }^{(2,2)} &=&0,\;a_{1,\gamma }^{(2,2)}=0,\;a_{2,\gamma
}^{(2,2)}=-\frac{1}{1+\gamma },  \nonumber \\
\delta _{\gamma }^{(2,3)} &=&-4\sqrt{\frac{4\gamma +7}{\left( 2\gamma
+3\right) \left( 2\gamma +5\right) }},\;a_{1,\gamma }^{(2,3)}=-2\sqrt{\frac{%
4\gamma +7}{\left( 2\gamma +3\right) \left( 2\gamma +5\right) }},\;
\nonumber \\
a_{2,\gamma }^{(2,3)} &=&\frac{2}{2\gamma +5}.  \label{eq:delta_a_j_n=2}
\end{eqnarray}
In this case $u_{\gamma }^{(2,1)}(x)$, $u_{\gamma }^{(2,2)}(x)$ and $%
u_{\gamma }^{(2,3)}(x)$ have zero, one and two nodes, respectively, in the
interval $0<x<\infty $. It is convenient to arrange the roots of $a_{n+1}=0$
as $\delta _{\gamma }^{(n,i)}>\delta _{\gamma }^{(n,i+1)}$ so that $%
u_{\gamma }^{(n,i)}(x)$ has $i-1$ nodes in $0<x<\infty $.

From the roots of $a_{2}=0$ the authors derive the frequency
\begin{equation}
\omega _{1,l}=\frac{4q^{2}}{m\left( bg\right) ^{2}\left( 2|\nu |+3\right) },
\label{eq:omega^(1)_OBB}
\end{equation}
and state that ``Hence, Eq. (26) (present equation (\ref{eq:omega^(1)_OBB}))
corresponds to the allowed values of the parameter $\omega $ that yield a
polynomial of first degree to the function $f(x)$. These are the allowed
values of $\omega $ with respect to the radial mode $n=1$. For this reason,
we have labelled $\omega =\omega _{1,l}$. For other radial modes ($%
n=2,3,4,\ldots $), other expressions can be obtained. Therefore, from Eq.
(26), we observe that there is a discrete set of values of the parameter $%
\omega $ that permit us to construct a polynomial of first degree to $f(x)$%
.'' They also show an analytical expression for $\omega _{2,l}$ and for the
energies $\mathcal{E}_{1,l,k}$ and $\mathcal{E}_{2,l,k}$ and state that
``The allowed energies (28) show us that the effects of the violation of the
Lorentz symmetry modify the spectrum of energy for an electron that
interacts with the nonuniform electric field. In this case, the effects of
Lorentz symmetry violation determine the possible values of the angular
frequency (26) associated with the radial mode $n=1$, i.e. the values of $%
\omega $ that allow us to obtain a polynomial of degree $n=1$ to the
function $f(x)$.'' In what follows we will show that these statements are
nonsensical and have no physical significance because they are a product of
the wrong belief that the only bound states are those given by the
truncation condition.

In the general case one would obtain (in present, more general, notation)
something like

\begin{equation}
\omega _{\gamma }^{(n,i)}=\frac{\left[ \delta _{\gamma }^{(n,i)}\right]
^{2}q^{2}}{2m\left( bg\right) ^{2}},\;i=1,2,\ldots ,n+1,
\label{eq:omega^(n)}
\end{equation}
and (also in present notation)
\begin{equation}
\mathcal{E}_{\gamma }^{(n,i)}=\frac{\omega _{\gamma }^{(n,i)}\left( n+\gamma
+1\right) }{2}-\frac{m\left( bg\right) ^{2}}{8q^{2}}\left[ \omega _{\gamma
}^{(n,i)}\right] ^{2}+\frac{k^{2}}{2m}.  \label{eq:E^(n,i)}
\end{equation}
If we omit the solutions $\delta =0$ that appear for $n$ even then we have $%
(n-1)/2$ different energies for $n$ odd and $n/2$ for $n$ even. Such a
multiplicity of solutions was not taken into account by Oliveira et al\cite
{OBB20}. However, this fact is of no importance because the dependence of $%
\omega $ on the node number $n$ and on the quantum numbers is an artifact of
the truncation condition that has no physical significance. All these
results are meaningless because $\mathcal{E}_{\gamma }^{(n,i)}$ are
eigenvalues of the set of models $V_{\gamma }^{(n,i)}(x)$, $i=1,2,\ldots
,n+1 $ while $\mathcal{E}_{\gamma ^{\prime }}^{(n^{\prime },i^{\prime })}$
are eigenvalues of completely different models $V_{\gamma ^{\prime
}}^{(n^{\prime },i^{\prime })}(x)$, $i^{\prime }=1,2,\ldots ,n^{\prime }+1$.
In other words, what the authors exhibit as the spectrum of a given
quantum-mechanical system are, in fact, some particular eigenvalues of
several different models. In what follows we show how to obtain the true
energies of the quantum-mechanical model reduced to the eigenvalue equation (%
\ref{eq:eigen_eq}).

Given the model parameters $m$, $q$, $g$, $b$, $\omega $ and $\rho $ we
obtain a value of $\delta $ and solve equation (\ref{eq:eigen_eq}) for those
functions $u(x)$ that satisfy the bound-state condition (\ref
{eq:bound_states}). In this way we obtain the allowed eigenvalues $%
W_{j,\gamma }$, $j=0,1,\ldots $, $W_{j,\gamma }<W_{j+1,\gamma }$ and the
true allowed model energies
\begin{equation}
\mathcal{E}_{j,\gamma ,k}=\frac{W_{j,\gamma }}{4}\omega +\frac{k^{2}}{2m}.
\label{eq:E_j,gam,k_present}
\end{equation}
Notice that there are eigenvalues and eigenfunctions for any value of $%
\omega $ so that this frequency is by no means discrete. In order to
illustrate this point we apply the Rayleigh-Ritz variational method with the
non-orthogonal basis set $\left\{ \phi _{j}(x)=x^{\gamma +j}\exp \left( -%
\frac{x^{2}}{2}\right) ,\;j=0,1,\ldots \right\} $. As a first example, we
arbitrarily choose $\delta =\delta _{0}^{(1,1)}=\frac{2\sqrt{6}}{3}$; the
first eigenvalues are: $W_{0,0}=W_{0}^{(1)}=\frac{10}{3}$, $%
W_{1,0}=8.417789723$, $W_{2,0}=13.16139523$, $W_{3,0}=17.76476931$, $%
W_{4,0}=22.28609551$. Notice that the truncation condition only yields the
ground state and misses all the other eigenvalues. The second example is $%
\delta =\delta _{0}^{(1,2)}=-\frac{2\sqrt{6}}{3}$ and in this case we have $%
W_{0,0}=0.3822700980$, $W_{1,0}=W_{0}^{(1)}=\frac{10}{3}$, $%
W_{2,0}=6.589162760$, $W_{3,0}=9.984807649$, $W_{4,0}=13.46286362$. Clearly,
in this case the truncation condition yields the first excited state and
misses all the other eigenvalues. In this two examples we have chosen values
of $\delta $ that are given by the truncation condition. If we choose $%
\delta =1$, which is not a root of $a_{n+1}=0$ for any value of $n$, we have
$W_{0,0}=2.840687067$, $W_{1,0}=7.506478794$, $W_{2,0}=11.96275335$, $%
W_{3,0}=16.33275291$, $W_{4,0}=20.65232861$. None of these eigenvalues can
be derived from the truncation condition. We appreciate that these results
satisfy the Hellmann-Feynman theorem (\ref{eq:HFT}) because $W_{\nu
,0}\left( \delta _{0}^{(1,2)}\right) <W_{\nu ,0}\left( 1\right) <W_{\nu
,0}\left( \delta _{0}^{(1,2)}\right) $, where $\delta _{0}^{(1,2)}<1<\delta
_{0}^{(1,1)}$. We also realize that $\omega $ is not quantized and that
there are bound states for any given value of this parameter.

\begin{figure}[]
\begin{center}
\includegraphics[width=6cm]{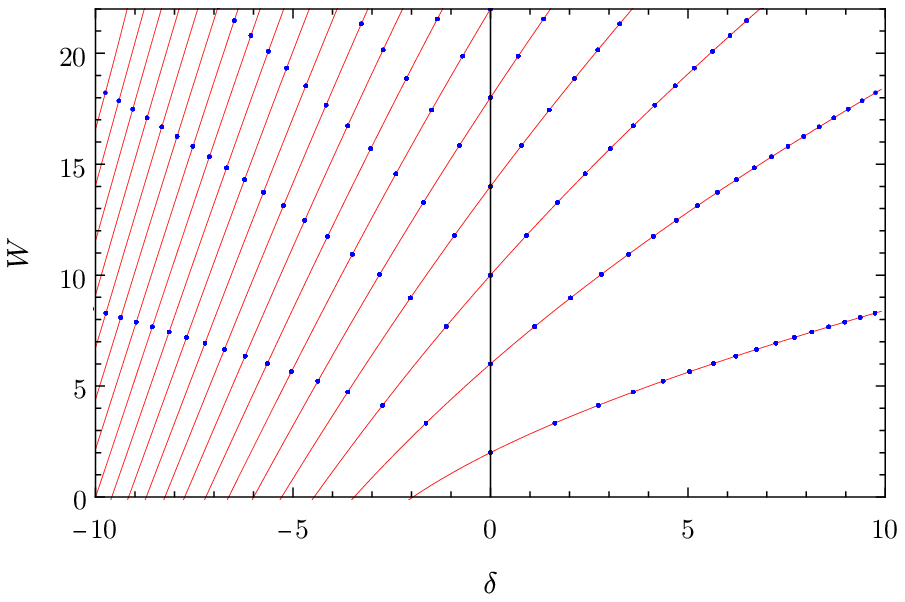} \includegraphics[width=6cm]{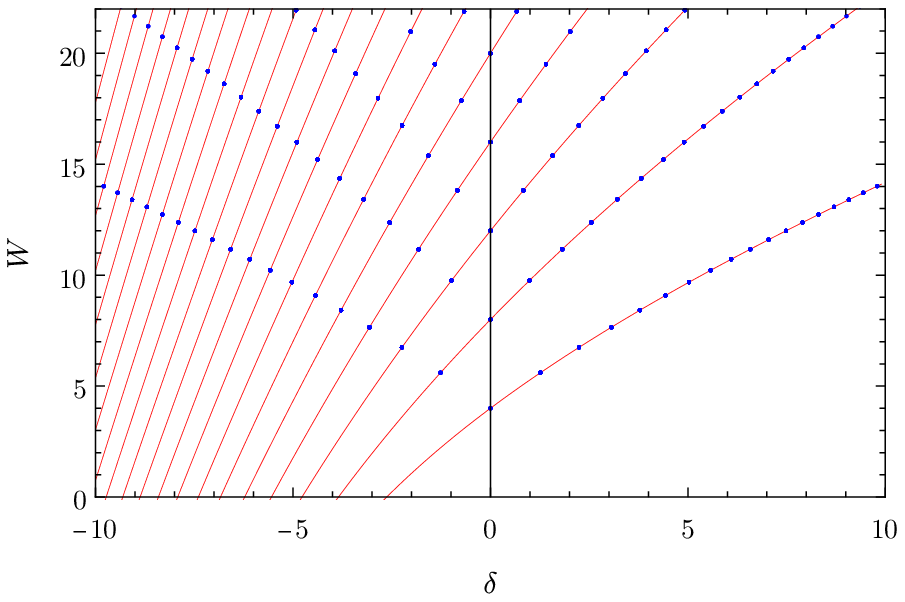}
\end{center}
\caption{Eigenvalues $W_\gamma^{(n,i)}$ from the truncation condition (blue
points) and variational results $W_{j,\gamma}$ (red lines) for $\gamma=0$
(left panel) and $\gamma=1$ (right panel)}
\label{Fig:Wjg}
\end{figure}

Figure~\ref{Fig:Wjg} shows several eigenvalues $W_{0}^{(n,i)}$ and $%
W_{1}^{(n,i)}$ given by the truncation condition (blue points) as well as
the eigenvalues $W_{j,0}$ and $W_{j,1}$ obtained by means of the variational
method (red lines). The continuous lines show that there are eigenvalues for
any value of $\delta $ and, consequently, for any value of $\omega $. The
occurrence of discrete values of the angular frequency is an artifact of the
truncation procedure that only yields some particular solutions to the
eigenvalue equation (\ref{eq:eigen_eq}) (red points). A question arises as
to the meaning of the roots $W_{\gamma }^{(n,i)}$ stemming from the
truncation condition. Taking into account the nodes of the solutions and the
Hellmann-Feynman theorem we conclude that a pair $\left( \delta _{\gamma
}^{(n,i)},W_{\gamma }^{(n,i)}\right) $ is a point on the curve $%
W_{i-1,\gamma }(\delta )$ in complete agreement to what is shown in figure~%
\ref{Fig:Wjg}.

\textit{Summarizing}: all the physical conclusions derived by Oliveira et al%
\cite{OBB20} regarding the effects of Lorentz symmetry violation that
determine the possible values of the angular frequency are obviously wrong
and a mere artifact of the truncation method used to obtain some particular
solutions to a conditionally solvable quantum-mechanical model (see, for
example,\cite{CDW00,AF20} and references therein).

\section*{Acknowledgements}

The research of P.A. was supported by Sistema Nacional de Investigadores
(M\'exico).

\end{document}